\documentclass[12pt, preprint]{aastex631}

\usepackage{lineno}

\begin{document}

\title{Multi-Passband Observations of A Solar Flare over  the \ion{He}{1} 10830~\AA\ line }

     \author{Yan Xu}
	 \affil{Institute for Space Weather Sciences, New Jersey Institute of Technology, 323 Martin Luther King Boulevard, Newark, NJ 07102-1982}
	 \affil{Big Bear Solar Observatory, New Jersey Institute of Technology, 40386 North Shore Lane, Big Bear City, CA 92314-9672, USA}
	
	 \author{Xu Yang}
	 \affil{Institute for Space Weather Sciences, New Jersey Institute of Technology, 323 Martin Luther King Boulevard, Newark, NJ 07102-1982}
	 \affil{Big Bear Solar Observatory, New Jersey Institute of Technology, 40386 North Shore Lane, Big Bear City, CA 92314-9672, USA}
	
    \author{Graham~S. Kerr}
	\affil{NASA Goddard Space Flight Center, Heliophysics Sciences Division, Code 671, 8800 Greenbelt Rd., Greenbelt, MD 20771, USA}
 	\affil{Department of Physics, Catholic University of America, 620 Michigan Avenue, Northeast, Washington, DC 20064, USA}

    \author{Vanessa Polito}
	\affiliation{Bay Area Environmental Research Institute, NASA Research Park,  Moffett Field, CA 94035-0001, USA}
    \affiliation{Lockheed Martin Solar and Astrophysics Laboratory, Building 252, 3251 Hanover Street, Palo Alto, CA 94304, USA}

    \author{Viacheslav~M. Sadykov}
	\affil{Physics \& Astronomy Department, Georgia State University, 25 Park Place NE, Atlanta, GA 30303, USA}

	 \author{Ju Jing}
	 \affil{Institute for Space Weather Sciences, New Jersey Institute of Technology, 323 Martin Luther King Boulevard, Newark, NJ 07102-1982}
	 \affil{Big Bear Solar Observatory, New Jersey Institute of Technology, 40386 North Shore Lane, Big Bear City, CA 92314-9672, USA}
	
	 \author{Wenda Cao}
	 \affil{Institute for Space Weather Sciences, New Jersey Institute of Technology, 323 Martin Luther King Boulevard, Newark, NJ 07102-1982}
	 \affil{Big Bear Solar Observatory, New Jersey Institute of Technology, 40386 North Shore Lane, Big Bear City, CA 92314-9672, USA}
	
	 \author{Haimin Wang}
	 \affil{Institute for Space Weather Sciences, New Jersey Institute of Technology, 323 Martin Luther King Boulevard, Newark, NJ 07102-1982}
	 \affil{Big Bear Solar Observatory, New Jersey Institute of Technology, 40386 North Shore Lane, Big Bear City, CA 92314-9672, USA}

\date{\today}
\begin{abstract}
This study presents a C3.0 flare observed by the BBSO/GST and IRIS, on 2018-May-28 around 17:10 UT. The Near Infrared Imaging Spectropolarimeter (NIRIS) of GST was set to spectral imaging mode to scan five spectral positions at $\pm 0.8$~\AA, $\pm 0.4$~\AA\ and line center of \ion{He}{1} 10830~\AA. At the flare ribbon's leading edge the line is observed to undergo enhanced absorption, while the rest of the ribbon is observed to be in emission.  When in emission, the contrast compared to the pre-flare ranges from about $30~\%$  to nearly $100~\%$ at different spectral positions. Two types of spectra, ``convex'' shape with higher intensity at line core and ``concave'' shape with higher emission in the line wings, are found at the trailing and peak flaring areas, respectively. On the ribbon front, negative contrasts, or enhanced absorption, of about $\sim 10\% - 20\%$ appear in all five  wavelengths. This observation  strongly  suggests that the negative flares observed in \ion{He}{1} 10830~\AA\ with mono-filtergram previously were not caused by pure Doppler shifts of this spectral line. Instead, \textbf{the enhanced absorption appears to be a consequence} of flare energy injection, namely non-thermal collisional ionization of helium caused by the precipitation of high energy electrons, as found in our recent numerical modeling results. In addition, though not strictly simultaneous, observations of \ion{Mg}{2} from the IRIS spacecraft, show an obvious central reversal pattern at the locations where enhanced absorption of \ion{He}{1} 10830~\AA\ is seen, which is in consistent with previous observations.

\end{abstract}

\clearpage

\section{Introduction}

The helium triplet around 10830 ~\AA\ are among the most important diagnostics of the chromosphere. Flare observations using \ion{He}{1} 10830~\AA\ line started several decades ago \citep{Harvey1984}, however, there have not been many events observed with this line since then \citep{You1992, Penn1995, Li2006, Zeng2014, Xu2016, Kobanov2018}. Most of the reported flares show emission similar to the observations in other chromospheric wavelengths, such as H$\alpha$. \citet{Li2007} studied spectrographic data of several flares and suspected that a certain level of nonthermal effects, such as those might occur in mid C-class flares, is needed for \ion{He}{1} 10830~\AA\ emission to reach a detectable level. With much higher spatial resolution, \citet{Zeng2014} reported emissions with more than $50\%$ contrast in a C3.9 flare observed by BBSO/GST.

In addition to the emission of the  \ion{He}{1} 10830~\AA\ line, \citet{Harvey1984} and \citet{Xu2016} presented enhanced absorption, which is one type of the ``negative'' flare originally defined in visible continuum \citep{Flesch1974, Hawley1995, Henoux1990, Ding2003b}. Previous studies have shown that enhanced absorption during flares were observed in another helium line, D3 at 5876~\AA\ \citep{Zirin1980, Liu2013}. It is worth noting that the shapes or sizes of negative flare ribbons are different. In the D3 negative flares, the flare sources are well defined ribbon-like structures and the absorption was enhanced on the entire ribbons \citep{Liu2013}. In the \ion{He}{1} 10830~\AA\ line, the darkening can be found in a large diffused area \citep{Harvey1984} or confined on the leading edge of the ribbon \citep{Xu2016}. Such differences may be caused by varying temperature and density in the flaring area or the band-width of the pre-filter.

There are two formation mechanisms of \ion{He}{1} 10830~\AA\ line, photoionisation- recombination mechanism (PRM) and collisional-recombination mechanism (CRM). The transition that forms \ion{He}{1} 10830~\AA\ line is in the triplet state of \ion{He}{1} (orthohelium), where the two electrons spin in the same direction of a helium atom. At the typical temperature and density of chromosphere, the transition from the ground state (parahelium), where the two electrons spin oppositely, to the orthohelium via direct collisional excitation is rare. Instead, in the quiet Sun, EUV or soft X-ray radiation can  populate orthohelium via PRM. Helium is ionised by coronal radiation, with recombination and cascades to orthohelium, generating an absorption feature \citep{Goldberg1939,Andretta1997,Centeno2008}. During a flare, non-thermal collisions between flare accelerated electrons and helium can introduce an alternative pathway to populating orthohelium. Helium can be ionised by these non-thermal collisions, with subsequent recombinations to orthohelium, which is the CRM mechanism. Determining which mechanism is responsible for the enhanced absorption and emission in flares is important in diagnosing flare energy transport mechanisms.

Recent numerical studies have shed light on the \ion{He}{1} 10830 \AA\ response to flares. \citet{Ding2005} used semi-empirical modelling to concludes that nonthermal CRM plays an important role in producing absorption at the initial phase of flares, though did not have a fully self-consistent model that included both PRM and nonthermal CRM. Using the F-CHROMA database (\url{https://star.pst.qub.ac.uk/wiki/public/solarmodels/start.html}), \citet{Huang2020} shows that the formation temperature and density of enhanced absorption are about $2\times10^{4}$~K and $6\times10^{11}$~cm$^{-3}$, respectively. Once the temperature or density increases above the thresholds, the \ion{He}{1} 10830~\AA\ line turns from absorption to emission. Their results are consistent with the prediction made by \citet{Zirin1980}. More recently, \citet{Kerr2021} presented detailed analysis of \ion{He}{1} 10830~\AA\ absorption during flares using RADYN \citep{Carlsson1992,Carlsson1997,Allred2015} simulations that included both PRM and nonthermal CRM. The authors confirmed that nonthermal CRM plays the key role in producing the darkening at the beginning of the flare. In their simulations omitting non-thermal collisions meant that the line did not undergo a period of enhanced absorption. In addition, they found that the level of darkening is related to the properties of the flare acceleraetd electrons, with a positive correlation between the low-energy cutoff of the precipitating electron beam. The `harder' the beam (that is, the greater the number of deeply penetrating higher energy electrons in the distribution), the stronger the absorption.

In the previous monochromatic narrow-band observations, the Lyot filter was set to the blue wing of the \ion{He}{1} line at $10830.05$ with a bandpass of 0.5~\AA\ \citep{Zeng2014, Xu2016}. It is representative of the line's general behaviour if the line profile remains symmetric or the intensity varies in the same direction for the entire line. However, one can imagine scenarios in which a part of the line is in emission while the rest remains absorption (e.g. due to Doppler motions). So, observations at a single, narrow, spectral position may not represent the characteristic of the entire line profile. Such a special case exists in some modeling results, such as the red profile (at t = 10 s) in Figure 1 panel (a) in \citet{Huang2020}. To confirm our previous conclusions that \ion{He}{1} 10830 \AA\ undergoes enhanced absorption at the flare ribbon's narrow leading edge and that those observations were not a serendipitous observation of an unusual line profile, we aimed to observe the 10830~\AA\ line at high spatial resolution and with rough spectral resolution. Such observations will also facilitate a more detailed model-data comparison. In this study, we present a C-3.0 flare, observed at five different spectral positions around the \ion{He}{1} 10830~\AA\ line, using the 1.6-m Goode Solar Telescope \citep[GST;][]{GST} at Big Bear Solar Observatory (BBSO). The flare was also observed by the Interface Region Imaging Spectrograph \citep[IRIS;][]{IRIS}.

\section{Observations}

\begin{figure}[ht]
\center
\includegraphics[scale=1.2]{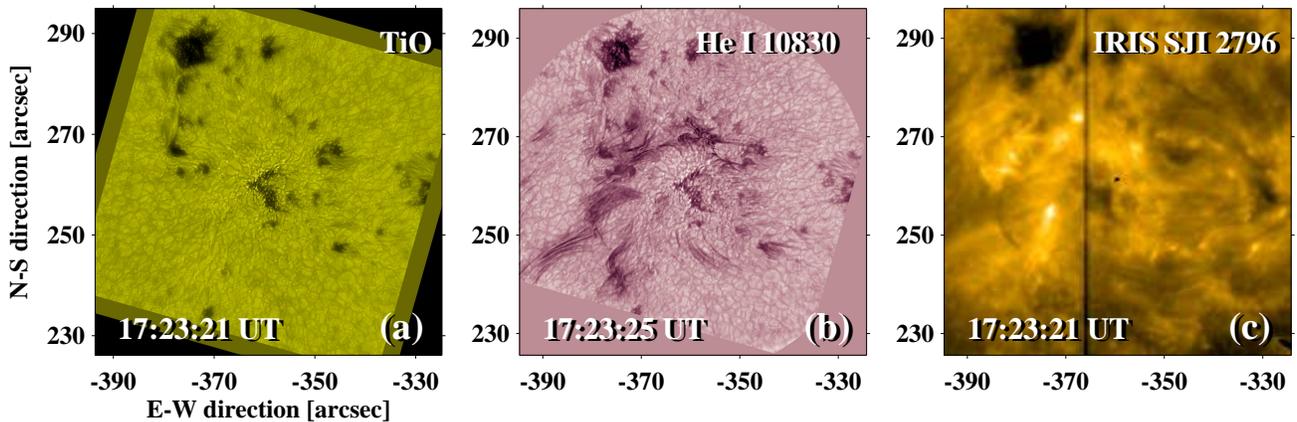}
\caption{Sample images taken around 17:23 UT for the C3.0 flare that occurred on 2018-May-28. Panel (a): TiO continuum image.
Panel (b): \ion{He}{1} 10830 red wing $+0.8$~\AA.
Panel (c): IRIS SJI 2796 image. }   \label{images}
\end{figure}

A C3.0 flare that occurred in active region (AR) 12127 on 2018 May, 28, peaking around 17:10~UT, was observed by both BBSO/GST and IRIS. BBSO/GST observations were carried out at three different channels, the  Broad-Band Filter Imager (BFI), Visible Imaging Spectrometer (VIS) and Near Infra-Red Imaging Spectropolarimeter \citep[NIRIS;][]{BBSONIRIS}, for TiO, H$\alpha$ and \ion{He}{1} 10830~\AA, respectively. NIRIS was set to the imaging mode without polarization for observations of the \ion{He}{1} 10830~\AA\ line at five spectral positions  (line center, $\pm 0.4$~\AA\ and $\pm 0.8$~\AA). The spectral calibrations of these two lines are done every day in quiet Sun areas at the beginning of the observations.  Multiple flats fields are taken for 10830 to compensate for minor temporal changes of the flat fields. The image scale is about 0.\arcsec062/pix and the effective cadence for a scan of five spectral positions is 45 s. Figure~\ref{images} shows sample images taken in these 3 channels and the IRIS slit-jaw imager (SJI) 2796\AA\ filter after the flare. The GST images are aligned with SDO/HMI continuum maps and displayed with heliocentric coordinates.

\section{Results}
This flare shows a circular shape of emission observed in the IRIS SJI UV bands, though GST's smaller field of view only caught a portion of the flare ribbons. There are ribbon-like features under the null point, and we concentrate on the southern ribbon in this study. Other ribbons or foot points are blocked or contaminated by the dark loops of the possible dome structure and highly dynamic fibrils. Because of the short duration, seeing variation, and complex structures, only a few locations with good images, between 17:09 and 17:11 UT, in \ion{He}{1} 10830 are suitable for analysis. Time sequence studies, as in \citet{Xu2016}, are not possible with this dataset. We present below a multi-wavelength analysis at a specific time, near the flare peak around 17:10 UT, for emission (positive contrast) and enhanced absorption (negative contrast) separately.

\subsection{Emission}

\begin{figure}[ht]
\center
\includegraphics[scale=1.2]{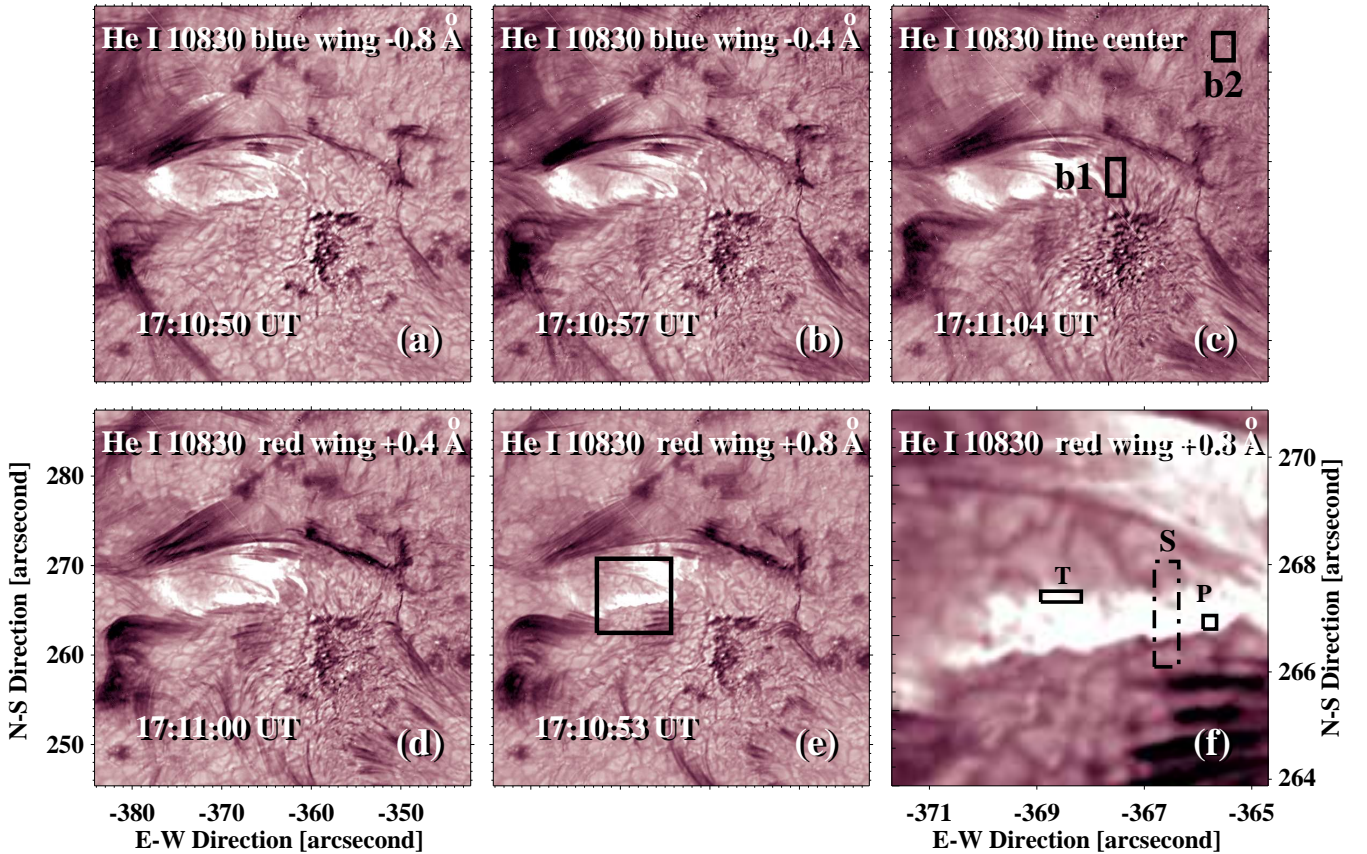}
\caption{\ion{He}{1} 10830 images taken by BBSO/GST at five spectral positions,  $\pm 0.8$~\AA, $\pm 0.4$~\AA\ and line center, in panels (a) to (e). Two areas, marked b1 and b2 in panel (c), are selected as the pre-flare background. The actual measurements of background intensities are obtained from images around 17:00:57 UT. Boxes of b1 and b2 are shown during the flare time for illustrative purpose only. In panel (e), the target ribbon is outlined by a black box. The small FOV within the black box is displayed in panel (f) for a closer view. The image scale is 0.\arcsec062/pix for all panels. Two small boxes in panel (f) indicate two flaring areas. The one close to the ribbon front represents the initial peak of the flare emission, marked as ``P''. The rectangular box near the center of panel (f) locates near the back edge of the flare ribbon, representing the trailing or well developed area of emission marked as ``T''. The dotted-dashed box marked as ``S'' is 10 pixels wide and shows the locations of 10 slits with the lower end covering the ribbon front with negative contrasts. The labels of panel (d) are shared by panels (a) to (e).  Panel (f) uses different labels.}   \label{spect}
\end{figure}

\begin{figure}[ht]
\center
\includegraphics[clip,width=0.8\textwidth]{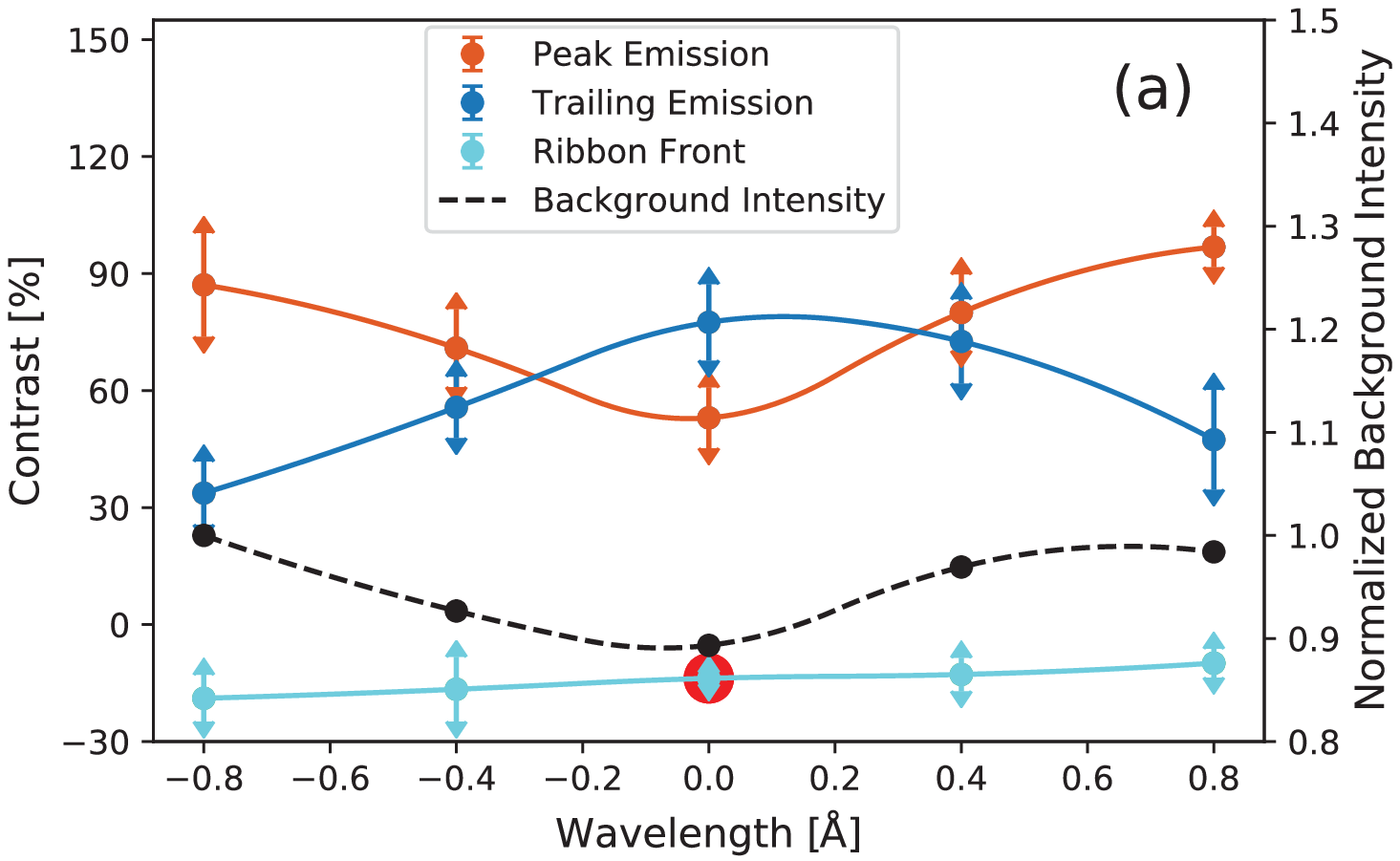}
\includegraphics[clip,width=0.73\textwidth]{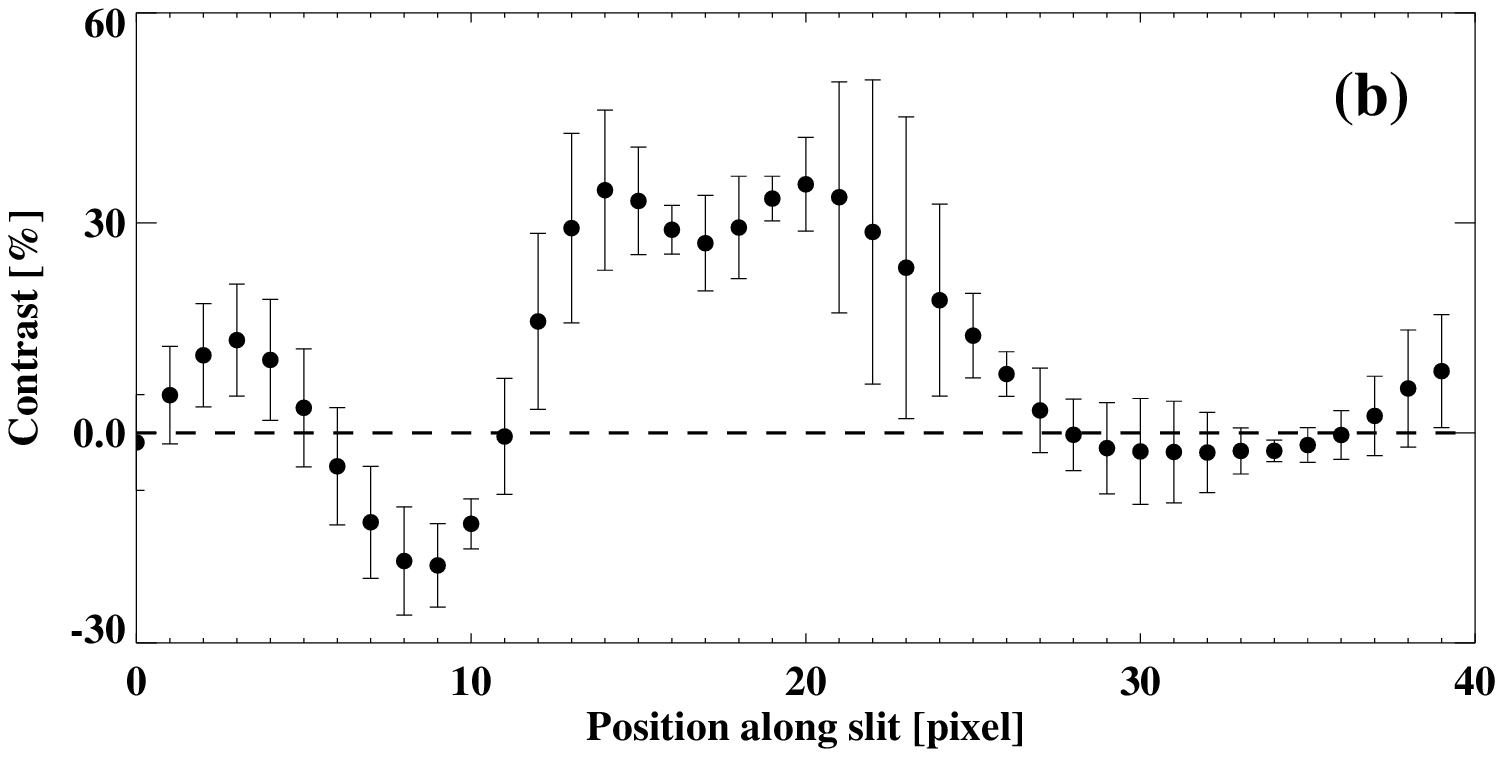}
\caption{\textbf{Panel (a)} shows pseudo line profiles of the \ion{He}{1} 10830~\AA\ line in different spatial position on the flare ribbon, with different colors. Orange: Emission profile in the peak flaring area (``P'' in panel (f) of Figure~\ref{spect}). Blue: Emission profile in the trailing area (``T'' in panel (f) of Figure~\ref{spect}). Cyan: {\it Absorption} profile on the ribbon front. The point of the line center is measured at a different location and therefore highlighted with a red ring. The 1-$\sigma$ uncertainty is calculated from the standard deviation and affected by the area selection. For example, larger areas tend to have larger uncertainties. All colored curves share the same Y-axis on the left side with a unit of percentage of contrast. The black dashed curve links the intensities of the background $b$, using the right Y-axis. The intensities are normalized to the maximum measured at -0.8~\AA\ and the absolute amplitudes are magnified by a factor of 10 to illustrate the differences among the five spectral positions.
\textbf{Panel (b)} shows the averaged contrasts along the 10 slits (dotted-dashed rectangular box marked as ``S'' in panel (f) of Figure~\ref{spect}), at -0.8~\AA.} \label{profiles}
\end{figure}

A representative set of images in each of the five spectral positions taken from 17:10:50 UT to 17:11:04 UT, at $\pm 0.8$~\AA, $\pm 0.4$~\AA\ and line center of \ion{He}{1} 10830~\AA\ line, are shown in Figure~\ref{spect} panels (a) to (e). The core region covering the target ribbon is outlined by a black box in panel (e) and the small FOV within the box is presented in panel (f). In front of the bright ribbon, there is a faint dark edge, which will be discussed in the next section concerning \ion{He}{1} 10830\AA\ absorption. Two different areas of emission are selected, one immediately behind the ribbon front and one in the trailing edge. They are marked by black box named ``T'' and ``P'' in panel (f), for trailing component and peak area, respectively. Image contrasts are defined as $contrast = (I - b)/b$, in which $I$ is the intensity and $b$ is the background intensity. The flaring area is surrounded by several small sunspots, groups of pores and dark dynamic fibrils. Background selection should not be based solely on the corners of the FOV, because the intensities in the corners are much lower than the center due to dominant dark features and relatively weak correction of the AO system. Therefore, we combine two areas, one in the center within a granulation area (b1 in  panel (c) of Figure~\ref{spect}) and one from the upper right corner outside the small pore (b2 in panel (c) of Figure~\ref{spect}). They are illustrated on the image at flare time, the actual measurement of the background intensities are carried out on the pre-flare frames around 17:00:57 UT. The black dots, connected by a dashed curve, in Figure~\ref{profiles} panel (a), shows the normalized (to the maximum intensity at -0.8~\AA, referring to the right Y-axis) intensity of the background, $b$. The intensities have been magnified by a factor of 10 to show the difference among the five spectral positions. In principle, the pseudo-line profile of the background $b$ is a very shallow absorption curve.

The contrasts of emission at five different wavelengths are listed in Table~\ref{tb} and plotted in panel (a) of  Figure~\ref{profiles} in blue (for the trailing area) and red (for the peak area) colors, and connected by a quadratic function, which is used for illustration purposes only. The standard deviation of contrasts of all pixels in box ``T'' and ``P'' is calculated and shown as one sigma uncertainties (arrows). The results show that the contrast enhancement ranges from about $30\%$ to nearly $100\%$ compared to the quiet Sun areas. These values compare well with the modeling results, which reached peak contrasts in excess of $150\%$ to $200\%$, depending on flare strength \citep[e.g.][]{Huang2020, Kerr2021}. The quadratic curves link the different spectral positions and work as pseudo line profiles, which can be compared with modeled spectra qualitatively. The peak area is close to the propagating ribbon front, so that it represents the early stage of the emission, following flare energy deposition (which in the standard flare model is primarily via bombardment of nonthermal electrons). The trailing area is away from the ribbon front and represents a well-developed stage of the flare, where thermal processes are presumed to dominate. Although the seeing conditions are not good enough for a detailed analysis of the temporal behaviour, we are able to compare our observations of different spatial features that represent different stages during the flare, with the temporal behavior of the \ion{He}{1} 10830~\AA\ line in flare simulations. In their Figure 1, panel (a), \citet{Huang2020} presented two types of \ion{He}{1} 10830 emission profiles, a center-reversed (at t = 2.9 s and 4.3 s) and a red-wing enhanced (at t=10 s). In addition, \citet{Kerr2021} also produced red-wing enhanced spectra (see their Figures 6 and 7). Our pseudo profiles show similar behavior. Due to the lack of spectral resolution in 10830 observations, the two types of observed spectra are named ``convex'' and ``concave'', representing higher emission in line core or line wing, respectively. The ``convex'' spectrum is shown in the trailing area and the ``concave'' spectrum dominates the peak area.

\subsection{Enhanced Absorption}

\citet{Xu2016} presented a dark ribbon front in the \ion{He}{1} 10830 blue wing ($10830.05 \pm 0.25$~\AA). In the multi-spectral-position observation under investigation, similar edges in front of the ribbon with negative contrasts are noticed in all of the five wavelengths channels. The area of 10 slits is highlighted by a dotted-dashed box ``S'' in Figure~\ref{spect} panel (f).  The contrasts along each slit are obtained against the same background area mentioned in the previous section. The average contrasts of the 10 slits are plotted in Figure~\ref{profiles} panel (b). The range of these 10 values at each point along the X-axis is used to evaluate the uncertainty. For instance, at the 9th point, the contrast level varies from $-13\%$ to $-25\%$. Therefore, the average contrast is as low as $-20\%$ with uncertainties of $\pm 6.0\%$. The same measurements are carried out for the other spectral positions and the results are plotted in Figure~\ref{profiles} panel (a) in cyan color and listed in Table~\ref{tb}. To avoid contamination from other dark features, such as fibrils, slit positions are slightly shifted (by 2 or 3 pixels) for offbands and significantly differ for the line center. Since the negative contrasts are not measured at a strictly identical location (as we do for the emission), pseudo line profiles of the negative contrasts are less meaningful compared with the emission profiles. The key information here is that the negative contrasts appeared in front of the bright flare ribbon at a level similar to the results in \citet{Xu2016}, \emph{and in all of the five wavelengths channels}. This result suggests that the enhanced absorption on the ribbon front is not caused by the minor absorption embedded in emission profiles, but that the full line profile is undergoing enhanced absorption.

\begin{table}[pht]
\caption{Contrasts [\%] at five different spectral positions. \label{tb}}
\vspace{-1em}
\centering
\begin{tabular}{lccccc}
\\
\tableline\tableline
Spectral Position    & -0.8~\AA      & -0.4~\AA      & Center           & +0.4~\AA     & +0.8~\AA     \\
\tableline

Peak Area (``P'')     & 87 $\pm$ 13  & 71 $\pm$ 10  & 53 $\pm$ 8          & 80 $\pm$ 10   & 97 $\pm$ 5  \\
Trailing Area (``T'') & 34 $\pm$ 8   & 56 $\pm$ 8   & 78 $\pm$ 10          & 73 $\pm$ 11  & 47 $\pm$ 13  \\
Front Edge            &-19 $\pm$ 6   & -17 $\pm$ 8  & -14 $\pm$ 4$^\star$ & -13 $\pm$ 4  & -10 $\pm$ 4  \\

\tableline
\end{tabular}
\\
\vspace{1em}
\footnotesize Positive values in areas ``P'' and ``T'' indicate flare emissions and negative values on the front edge represent enhanced absorptions. ${}^\star$The slit positions for the line center measurements are quite different from those for the wings. Consequently, the negative contrast at line center is not comparable with the offbands.
\vspace{-1em}
\end{table}

\subsection{IRIS UV spectra}

\begin{figure}[hbt]
\center
\includegraphics[scale=1]{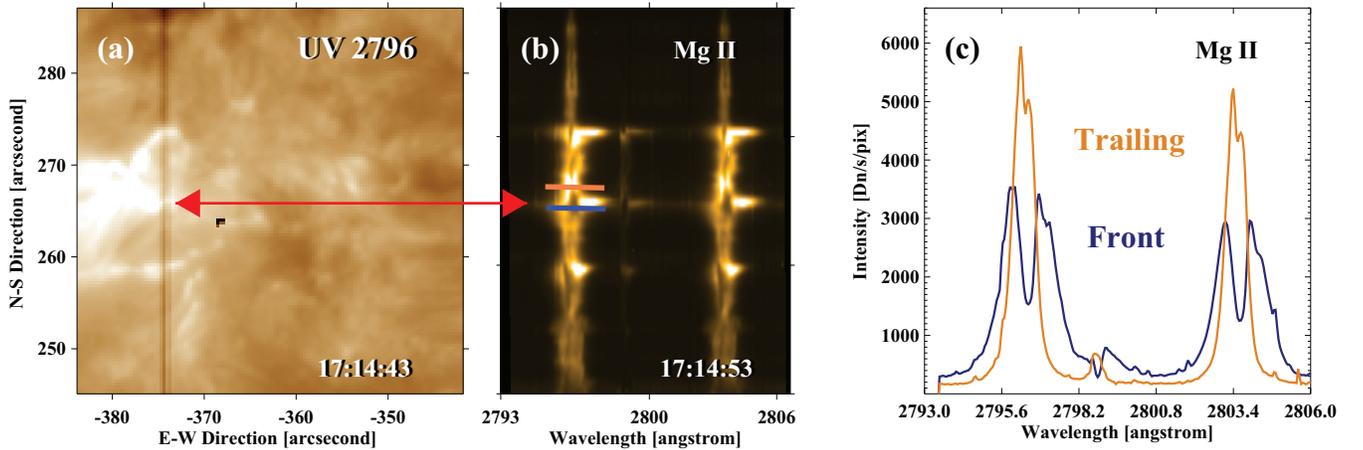}
\caption{IRIS SJI and spectra taken at 17:14 UT.
Panel (a): SJI 2796 image, in which the dark line indicates the location of slit.
Panel (b): \ion{Mg}{2} spectrum.
Panel (c): \ion{Mg}{2} K line profiles for the northern flare ribbon (Positive Flare), the southern ribbon where enhanced \ion{He}{1} 10830 is located (Negative Flare), and the flaring loops at bottom (Flare Loop). The positions of spectra are marked using the same color in panel (b).  }   \label{IRIS}
\end{figure}

IRIS also observed AR 12712 on 2018-May-28, with a very large 320-step raster. Figure~\ref{IRIS} shows a UV 2796~\AA\ IRIS SJI image and a \ion{Mg}{2} spectrum. This data set was taken around 17:14 UT, which is the earliest time at which the spectrograph slit has a clear view of the flare ribbon (previously, the slit's view of the ribbon was obscured by the dome structure). By comparing the locations of the flare ribbons on the IRIS SJI and \ion{He}{1} 10830 images, we could identify the ribbon location in the SJI and \ion{Mg}{2} spectral images, as pointed out by the red arrow in panels (a) and (b) of Figure~\ref{IRIS}. The ribbon propagates from north to south, so that the ribbon front is located in the southern edge. Note that here we aim to compare behaviour of \ion{Mg}{2} h \& k lines at the ribbon front, which has propagated slightly south of where we identified the \ion{He}{1} 10830~\AA\ ribbon edge at the earlier time, to those profiles in the trailing portion of the ribbon. The \ion{Mg}{2} spectrum at $Y=239$~pixel indicated by the blue bar in panel (b),which is a representative position of the ribbon front, is plotted using blue color in panel (c). We also note that the \ion{He}{1} 10830~\AA\ negative contrasts were present at the ribbon front also. The \ion{Mg}{2} lines  at the ribbon front are characterized by a significant central reversal and line broadening. A sample spectrum in the trailing area is selected eight pixels north to the ribbon front ($Y = 247$~pixel, indicated by the orange bar in panel (b)) and plotted using orange color. The ribbon-trailing spectrum is located within the well-developed flaring area with much larger intensities than the spectrum at the ribbon front. However, in such spectrum the line broadening is reduced and the center reversal is barely seen. The spectral properties and differences at the ribbon front and trailing area found in this study agrees with previous observations \citep{Xu2016, Panos2018}. Modelling of the \ion{Mg}{2} spectra alongside the \ion{He}{1} spectra is currently underway to see if we can reproduce these patterns (Kerr et al, \textsl{in prep}).

\section{Discussion}

In this study, we present the analysis of a C3.0 flare observed on May 28, 2018, at five spectral positions around the \ion{He}{1} 10830~\AA\ line. The emission level of $30\% \sim100\%$ contrast and the two types of line profiles (``convex'' and ``concave'') agree well with previous modeling results \citep{Huang2020, Kerr2021}. Other kinds of spectra at different stages in the flare evolution likely exist and should be studied with future observations under better seeing condition and/or better spectral resolutions. Enhanced absorption is detected in front of the moving flare ribbon, which is at a similar level of that reported in \citet{Xu2016}. The blue wing tends to have stronger enhanced absorption than the red wing, which offers a good feature for comparison to flare models of this line. There is a time gap of  $\sim200$~s between the  \ion{Mg}{2} spectra shown in Figure~\ref{IRIS} and the \ion{He}{1} 10830~\AA\ images shown in Figure~\ref{spect}, which are associated with two close, but distinct emission peaks in the GOES soft X-ray emission. By inspecting the IRIS SJI images at the times corresponding to the two different peaks, we find that the morphology of the flare emission evolves significantly only in the region located in the left part of the flaring area. The propagation direction and shape of the target flare ribbon near the center of the flaring area did not change substantially. Future observations, ideally with improved temporal and  spectral resolutions, should be sought for this purpose. Our results do confirm that single passband observations are suitable for studying the enhanced absorption feature; if a choice must be made, the blue wing offers a cleaner view of the negative contrast.

\begin{acknowledgements}\par
We would like to thank the referee for valuable comments. This work is supported by NSF under grants AGS 1821294, 1954737, 1936361, and AST 2108235, and by the NASA under grants 80NSSC17K0016, 80NSSC19K0257, 80NSSC19K0859, 80NSSC21K1671, 80NSSC20K0716, and 80NSSC21K0003. The BBSO operation is supported by NJIT, US NSF AGS 1821294, and the GST operation is partly supported by the Korea Astronomy, Space Science Institute, Seoul National University. The supportive white light data is taken by SDO/HMI. IRIS is a NASA Small Explorer mission developed and operated by LMSAL with mission operations executed at NASA Ames Research center and major contributions to downlink communications funded by the Norwegian Space Center (NSC,Norway) through an ESA PRODEX contract.
\end{acknowledgements}


\end{document}